\documentclass[11pt,letterpaper]{article}
\usepackage{color}
\usepackage{amsmath}
\usepackage{amsthm}
\usepackage{amsfonts}
\usepackage{amssymb}
\usepackage{graphicx}
\usepackage{float}
\usepackage{geometry}
\usepackage[utf8]{inputenc}

\usepackage[all]{xy}
\usepackage{amscd}
\usepackage{amsthm}
\usepackage{longtable}
\usepackage{mathrsfs}
\usepackage{amsmath}
\usepackage{amssymb}
\usepackage{enumitem}
\usepackage[table,dvipsnames]{xcolor}

\usepackage{soul}

\usepackage{booktabs}

\usepackage{array}

\usepackage{pdflscape}
\usepackage{listings}
\lstdefinestyle{myCustomRStyle}{
  language=R,
  numbersep=10pt,
  basicstyle=\ttfamily\tiny,
  tabsize=4,
  showspaces=false,
  showstringspaces=false,
  backgroundcolor = \color{white},
  rulecolor=\color{white},
  breakatwhitespace=false,
  breaklines=true,
  keepspaces=true,
  showspaces=false,
  showstringspaces=false,
  showtabs=false
}
\usepackage{xcolor}
\usepackage{colortbl}

\definecolor{Gray}{gray}{0.80}

\usepackage{color}

\definecolor{dkgreen}{rgb}{0,0.6,0}
\definecolor{gray}{rgb}{0.5,0.5,0.5}
\definecolor{mauve}{rgb}{0.58,0,0.82}

\usepackage{caption}

\usepackage[utf8]{inputenc}
\usepackage[english]{babel}

\usepackage[thinlines]{easytable}

\usepackage{tikz}
\tikzset{> = stealth} 
\usetikzlibrary{positioning}

\usepackage{pdfpages}

\usepackage{listings} 
\usepackage{xcolor}

\usepackage{comment}
\usepackage{color}
\usepackage[backref=page]{hyperref}
\hypersetup{
 colorlinks=true,
 linkcolor=blue,
 filecolor=magenta,
 citecolor=darkRed,
 urlcolor=cyan,
 pdftitle={
 Individual causal effect estimation accounting for latent disease state modification among bipolar participants in mobile health studies},
 }

\definecolor{codegreen}{rgb}{0,0.6,0}
\definecolor{codegray}{rgb}{0.5,0.5,0.5}
\definecolor{codepurple}{rgb}{0.58,0,0.82}
\definecolor{backcolour}{rgb}{0.95,0.95,0.92}
\definecolor{darkRed}{rgb}{0.60,.03,.03}

\lstdefinestyle{mystyle}{
    backgroundcolor=\color{backcolour},   
    commentstyle=\color{codegreen},
    keywordstyle=\color{magenta},
    numberstyle=\tiny\color{codegray},
    stringstyle=\color{codepurple},
    basicstyle=\ttfamily\footnotesize,
    breakatwhitespace=false,         
    breaklines=true,                 
    captionpos=b,                    
    keepspaces=true,                 
    numbersep=5pt,                  
    showspaces=false,                
    showstringspaces=false,
    showtabs=false,                  
    tabsize=2
}

\theoremstyle{definition}

\newtheorem*{counter example}{Counter Example}

\DeclareUnicodeCharacter{2008}{\textcolor{red}{BAD!!}}

\begin{document}

\def\spacingset#1{\renewcommand{\baselinestretch}{#1}\small\normalsize} \spacingset{1}

\title{{\Large Individual causal effect estimation accounting for latent disease state modification among bipolar participants in mobile health studies}}
\author{Charlotte R. Fowler$^1$, Xiaoxuan Cai$^2$, Habiballah Rahimi-Eichi$^3$, Lisa Dixon$^4$, \\ Justin T. Baker$^3$, Jukka-Pekka Onnela$^5$, Linda Valeri$^6$}
\date{}

\clearpage \maketitle


\vspace{-10mm}

$^1$ Department of Mathematical Sciences, Worcester Polytechnic Institute, Worcester, MA, 01609, USA

$^2$ Department of Statistics, Ohio State University, Columbus, OH, 43210, USA

$^3$ Institute for Technology in Psychiatry, McLean Hospital, Harvard Medical School, Belmont, MA, 02478, USA

$^4$ New York State Psychiatric Institute and Department of Psychiatry, Columbia University, New York, NY, 10032, USA

$^5$ Department of Biostatistics, Harvard T.H. Chan School of Public Health, Boston, MA, 02115, USA

$^6$ Department of Biostatistics, Columbia University, New York, NY, 10032, USA

\begin{abstract}

Individuals with bipolar disorder tend to cycle through disease states such as depression and mania. The heterogeneous nature of disease across states complicates the evaluation of interventions for bipolar disorder patients, as varied interventional success is observed within and across individuals. In fact, we hypothesize that disease state acts as an effect modifier for the causal effect of a given intervention on health outcomes. To address this dilemma, we propose an N-of-1 approach using an adapted autoregressive hidden Markov model, applied to longitudinal mobile health data collected from individuals with bipolar disorder. This method allows us to identify a latent variable from mobile health data to be treated as an effect modifier between the exposure and outcome of interest while allowing for missing data in the outcome. A counterfactual approach is employed for causal inference and to obtain a g-formula estimator to recover said effect. The performance of the proposed method is compared with a naive approach across extensive simulations and application to a multi-year smartphone study of bipolar patients, evaluating the individual effect of digital social activity on sleep duration across different latent disease states.

\end{abstract}

\newpage 
\setcounter{page}{1}
Severe mental illness serves as a class to highlight the most serious and long-lasting mental illness conditions, including schizophrenia, severe bipolar disorder, and severe major depression. Individuals living with severe mental illness (SMI) often face difficulties accessing care, and have poorer health outcomes and reduced life expectancy \cite{drake2014recovery, jayatilleke2017contributions}. However, the treatment of these individuals is complicated by the heterogeneous nature of SMI between patients and disease states, resulting in varied interventional success \cite{mills2021detecting, volavka2008heterogeneity, mohr2004heterogeneity}. Heterogeneity arises not only from differences in the effect of a given intervention between individuals, but also from differences in the disease states one experiences such as mania, depression, and euthymia. Even when two patients have the same diagnosis, this does not guarantee that they both experience the same disease states or that their disease states manifest in similar ways \cite{angst1978course, angst2004bipolar}. This heterogeneity necessitates an individualized analysis approach that can contend with complex causal relationships. 

Mobile health (mHealth) studies are a growing body of research that rely on digital technology to record exposures and outcomes in real time \cite{onnela2016harnessing, torous2016new}. This study design provides rich longitudinal data for each participant and can fill in informational gaps between in-person follow-up with patients. It is particularly useful to study severe mental illness (SMI), such as schizophrenia and bipolar disorder, as the mobile health approach allows researchers to capture fluctuations between disease states and details of at-home behavior that may not be evident from clinical visits \cite{torous2018characterizing}. Specifically, by employing mobile devices, we are able to gather the intensive longitudinal data needed to study potential at-home interventions at the individual level. 
Here, we consider the Bipolar Longitudinal Study, a cohort study from McLean Hospital which collected mobile health data from 74 patients with bipolar disorder, schizophrenia, or schizoaffective disorder for up to five years \cite{cai2022state, wang2021smartphone}. 

Our objective with mobile health data is to study causal effects between a given intervention and outcome, where the cause and effect change over time. Due to the heterogeneous nature of individuals living with SMI and the data collected, we employ an N-of-1 approach where participants and their relative causal effects are studied at the individual level, rather than the aggregate \cite{daza2018causal, cai2022state, cai2024causal}. 
However, we face several challenges in estimating the individual causal effect, including not directly observing the disease state and missing data. 
Disease state, representing a subject's current depressive, manic, or euthymic state, likely acts as an effect modifier of the causal effect of at-home interventions on daily health outcomes. Bipolar disorder is characterized by a patient exhibiting mood states of mania, hypomania, and depression \cite{grande2016bipolar, brady2017differential}. It has been shown that a disease state framework is also helpful to understand the mood and impairment faced by individuals with schizophrenia across time \cite{mohr2004heterogeneity, haro2018understanding}. 
In both populations, it is understood that disease state can generate heterogeneous intervention effects within an individual \cite{mohr2004heterogeneity, grande2016bipolar, van2020mood}, thus acting as a modifier. Accurately estimating the causal effect of a given intervention on health outcomes across disease states for individuals living with SMI will inform clinicians when giving recommendations to improve quality of life and disease prognosis. 
For example, when studying the effect of physical activity on sleep, the impact of physical activity on sleep might be different depending on the presence of a depressive or manic state thus modifying the observed effect. 
Unfortunately, it is difficult to observe an individual's disease state directly from mHealth data. 
Here, we assume that we collect sufficient information reflective of the disease state such as energy expenditure, mood, and symptoms, such that the state itself is latent. Under this hypothesis, to account for the modification by disease state, we first establish the latent disease state from mobile health data, and secondly include this state when estimating the individual causal effect of interest. 


There has been extensive work on identifying latent modifiers through latent modeling techniques. Specifically, this framework is common in the heterogeneous treatment effect literature, where a latent modifier is identified to explain the differences in observed effects between individuals in the sample \cite{suk2021hybridizing, lyu2023estimating, kim2015multilevel, kim2019estimating, shahn2017latent, post2022distribution, wang2021causal}. 
These latent modeling methods have been used to better understand heterogeneous effects in a range of settings including the impact of private tutoring \cite{suk2021hybridizing, lyu2023estimating}, pediatric kidney transplant effectiveness \cite{kim2015multilevel}, and the impact of Medicaid enrollment on emergency department utilization \cite{shahn2017latent}. 
However, these methods rely on the identification of latent subgroups among individuals in the sample. In the context of latent disease states modifying the individual causal effect, we instead aim to identify the latent classes explaining heterogeneous effects across time points for a given individual. Notably, when differences in the impact or direction of a given treatment exist and are undetected, we may spuriously conclude a treatment to be beneficial or harmful by aggregating across all time. Identifying such heterogeneity not only provides richer interpretation of results but allows for more accurate recommendations dependent on the individual's current disease state.   



To account for a latent variable representing disease state which acts as an effect modifier for the effect of interest, we employ a forward and backward algorithm to learn the disease state from our observed multivariate time series. 
The hidden Markov model (HMM) provides a framework to identify a latent state from its response time series through estimation of the initial state probabilities, a transition matrix, and the response distribution conditional on the latent state \cite{rabiner1986introduction, rabiner1989tutorial}. It is a natural choice to identify latent disease states from mHealth data given its design for time series data with a latent multinomial covariate. While an HMM relies on the Markov property of the response conditional on the latent state, we develop an approach that relaxes this assumption within an auto-regressive hidden Markov model framework \cite{juang1985mixture, ephraim2005revisiting}. Our proposed model also draws from literature on missing data in HMMs \cite{popov2016training, cooke2001robust, speekenbrink2021ignorable, dang2017learning}, and allows for missingness in the outcome through marginalization and recursive multiple imputation of lagged regressors. Notably, hidden Markov models have been employed previously in similar contexts. Hulme et al. employ continuous-time hidden Markov models to the ecological momentary assessment data of individuals with schizophrenia to predict transition to a latent alert state \cite{hulme2020adaptive}, and Ren \& Barnet use mixed effect hidden Markov modeling to study diurnal rhythms among adolescents with mood disorders \cite{ren2023combining}. However, the majority of literature for HMMs has the goal of latent state prediction or identification, not causal estimation, and typically involves less complex assumptions about variable relations \cite{eddy2004hidden, fine1998hierarchical, varga1990hidden}. Here, we expand on the existing hidden Markov model literature to our situation of interest and evaluate the performance when estimating an individual causal effect. Specifically, we develop an estimation strategy leveraging the counterfactual approach for causal inference and the HMM for estimation of the individual causal effect and inference. 

In the remainder of this paper, we will first define our causal estimand of interest and its relative g-formula. Next, summarize traditional notation and algorithms for constructing an auto-regressive hidden Markov model and explain our adaptations to allow us to fit the model to mobile health data. To evaluate our proposed methods, we compare them in simulations, in application to the Bipolar Longitudinal Study leveraging both actively and passively recorded data. Lastly, we discuss the strengths and limitations of our work, and the future directions needed.

\section{Causal framework}
\label{s:causal}

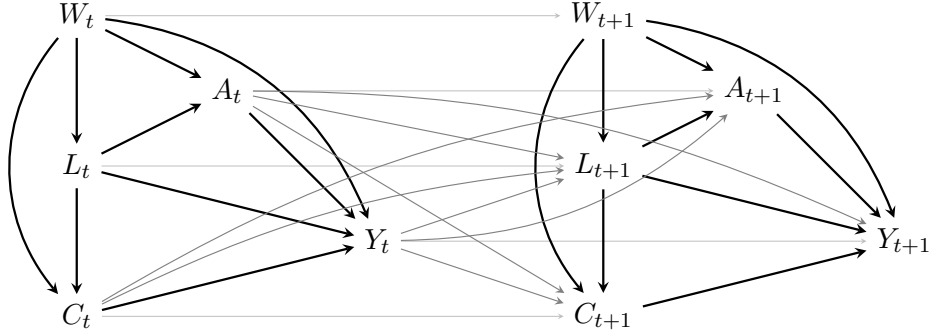
\begin{figure}
\begin{center}
\begin{tikzpicture}
\node (W1) at (0, 4) {$W_t$};
\node (L1) at (0, 2) {$L_t$};
\node (C1) at (0, 0) {$C_t$};
\node (A1) at (2,3)  {$A_t$};
\node (Y1) at (4,1)  {$Y_t$};
\node (W2) at (7, 4) {$W_{t+1}$};
\node (L2) at (7, 2) {$L_{t+1}$};
\node (C2) at (7, 0) {$C_{t+1}$};
\node (A2) at (9,3)  {$A_{t+1}$};
\node (Y2) at (11,1)  {$Y_{t+1}$};

\draw[->] [line width=0mm,   lightgray]
(W1) 	edge (W2)
(A1) 	edge (A2)
(C1) 	edge (C2)
(Y1) 	edge (Y2)
(L1) 	edge (L2);

\draw[->] [line width=.3mm,   black] 
(W1)	edge [bend right=40] (C1) 
(W1)	edge (L1)
(W1)	edge [bend left=30] (Y1)
(W1)	edge (A1)
(L1)	edge (A1)
(L1)	edge (Y1)
(L1)	edge (C1)
(C1)	edge (Y1)
(A1)	edge (Y1);
\draw[->] [line width=.3mm,   black] 
(W2)	edge [bend right=40] (C2) 
(W2)	edge (L2)
(W2)	edge [bend left=30] (Y2)
(W2)	edge (A2)
(L2)	edge (A2)
(L2)	edge (Y2)
(L2)	edge (C2)
(C2)	edge (Y2)
(A2)	edge (Y2);

\draw[->] [line width=.15mm,   gray] 
(A1)	edge [bend left=12] (Y2)
(A1)	edge (L2)
(A1)	edge (C2)
(C1)	edge [bend left=10] (L2)
(Y1)	edge (L2)
(Y1)	edge (C2)
(Y1)	edge [bend right=20] (A2)
(C1)	edge [bend left=12] (A2);


\end{tikzpicture}
\end{center}

\caption{Directed Acyclic Graph of hypothesized causal relationships between outcome $Y_t$, exposure  $A_t$, latent modifier $L_t$, observed confounders that are impacted by the latent state $C_t$, and observed confounders that operate independently of the latent state $W_t$. Black lines show contemporaneous causal relations, while grey lines show lagged causal relations.}
\label{fig p2 dag}
\end{figure}

We assume a follow-up length of $T$, and let $t$ index across days of follow-up, such that $t = 1,..., T$. We denote the outcome of interest as $Y_t$, exposure as $A_t$, observed confounders that are impacted by the latent state, such as reported symptoms, as $C_t$, and observed confounders that operate independently of the latent state, such as outside temperature, as $W_t$. Lastly, we let $L_t$ represent the latent disease state, with $J$ possible values. 
We maintain a flexible framework, where the outcome and/or exposure may be continuous or discrete. Our initial hypothesized causal structure for two time points is illustrated in a directed acyclic graph (DAG) in Figure \ref{fig p2 dag}. This framework could be easily adapted to incorporate additional observed confounders and lagged dependencies.
Notably, we impose a strong assumption by not including a directed edge between $A_t$ and $C_t$ as the directionality of such an edge is unknown and likely bi-directional throughout the day. For example, if $A_t$ represents daily texts and $C_t$ represents physical activity, it is possible that during a day both variables may affect each other. Thus to avoid imposing an incorrect directional assumption, we omit this edge and include lagged effects between the variables ($A_{t}$ to $C_{t+1}$, and $C_{t}$ to $A_{t+1}$).

\subsection{Contemporaneous Average Individual Causal Effect}

Our main focus is estimating the given effect of an intervention at time $t$ on the contemporaneous outcome measured within the same unit of time. To define said treatment effect of interest, we use potential outcome notation as developed by \cite{rubin1974estimating, splawa1990application, robins1989analysis, cai2024causal}. We first consider the scenario where we wish to interpret this effect for a given value of the latent disease state.
Specifically, we define the contemporaneous conditional average individual causal effect of $A_t = a_{1,t}$ versus $A_t = a_{0,t}$ on $Y_t$ given $L_t = j$ as  
\begin{align}
    E(Y_t^{a_{1,t}}- Y_t^{a_{0,t}}|Y_{t-1}, C_t, W_t, A_{t-1}, L_t = j).
\end{align}
We additionally wish to evaluate the contemporaneous effect without having conditioned on a given disease state. This would be of particular interest in the case where the disease state is unknown or not able to be clinically interpreted, for example within an online learning algorithm that provides recommendations in real-time \cite{trella2022designing}. We define the contemporaneous average individual causal effect of $A_t = a_{1,t}$ versus $A_t = a_{0,t}$ on $Y_t$ as
\begin{align}
    E(Y_t^{a_{1,t}}- Y_t^{a_{0,t}}|Y_{t-1}, C_t, W_t, A_{t-1}).
\end{align}

\noindent This effect is still to be calculated conditional of all observed covariates but will marginalize across the latent disease state. 

To estimate the contemporaneous causal effect of $A_t$ on $Y_t$, we require several key assumptions \cite{robins1994correcting, robins1997causal, boruvka2018assessing}. First, we assume consistency, meaning that $Y_t|A_t = a_t$ is equal to $Y^{a_t}$ for $t = 1,..,T$. This ensures that our observed data will be equal to the potential outcomes as defined above. Secondly, we assume positivity, i.e. that $P(A_t = a_t | A_{t-1}, Y_{t-1}, C_{t-1}, W_t) > 0$ for all observed $A_{t-1}, Y_{t-1}, C_{t-1}, W_t$. Lastly, we assume conditional exchangeability, or that $Y_t^{a_t} \perp\kern-5pt\perp A_t|Y_{t-1}, A_{t-1}, C_t, W_t, L_t$. 

\subsection{k-Lagged Average Individual Causal Effect}

As a secondary question, we wish to calculate the effect of a past intervention at time $t-k$ on a future outcome at time $t$ for some $k\geq 1$ \cite{boruvka2018assessing}. 
To denote this effect, we adopt notation from Boruvka et al. \cite{boruvka2018assessing}. 
For $k=1$ this effect will be comprised of direct and indirect effects, while for $k>1$ it will be entirely indirect. Thus, we wish to condition only on historical variables that are not descendants of the exposure, $A_{t-k}$. For brevity we will denote the observed regressors for $Y_t$ as $X_{Y,t} = (Y_{t-1}, A_t, A_{t-1}, C_t, W_t)$. We first consider the case where we condition on the value of the latent state at time $t-k$, the k-lagged conditional average individual causal effect, defined as: 

\begin{align}
E(Y_t^{a_{1,t-k}}- Y_t^{a_{0,t-k}}| L_{t-k} = j, X_{Y,t-k}\backslash A_{t-k}).
\end{align}

\noindent As with the contemporaneous causal effect, we may additionally be interested in the k-lagged effect without conditioning on the latent state at time $t-k$. This would be captured by the k-lagged average individual causal effect:

\begin{align}
E(Y_t^{a_{1,t-k}}- Y_t^{a_{0,t-k}}|X_{Y,t-k} \backslash A_{t-k}).
\end{align}

\noindent To estimate the k-lagged effect, we again need three common assumptions. Specifically, we assume consistency ($Y_t|A_{t-k} = a_{t-k}$ is equal to $Y^{a_{t-k}}$ for $t = 1,...,T$), positivity ($P(A_{t-k} = a_{t-k} | X_{A,t-k}) > 0$ for all $P(X_{A,t-k} = x_{A,t-k})>0$), and conditional exchangeability ($Y_t^{a_{t-k}} \perp\kern-5pt\perp A_{t-k}|X_{Y,t-k}, L_{t-k}$). We note that there may be additional interest in evaluating the effect of a sequence of interventions, not an intervention at a singular time point. A framework for this type of estimand has been explored by \cite{boruvka2018assessing}. 
While our work could be extended to answer such questions, it is beyond the scope of this paper. 


\section{Auto regressive hidden Markov model}
\label{s:HMM}






Hidden Markov models are composed of two variable types:   the observed response data ($Y_t$) and the latent categorical state $L_t$ \cite{rabiner1986introduction, rabiner1989tutorial}. In a traditional hidden Markov model, the response variable ($Y_t$) is caused only by the latent state ($L_t$), and the latent state ($L_t$) is caused only by the previous value of the state ($L_{t-1}$). Thus, one typically assumes that $L_t \perp\kern-5pt\perp L_{t-k}|L_{t-1}$ for any $k \geq 2$, and $Y_t \perp\kern-5pt\perp Y_{t-k} | L_t$ for any $k\geq 1$. Under these assumptions, there are three types of parameters to characterize the process. First, there is the initial latent state probability, $\pi_i = P(L_1 = i)$, for $i = 1,..., M$. Second, the relationship between the previous and current hidden state is characterized using a transition probability matrix $\textbf{A}$, where $a_{ij}$ is the probability of the current state $L_t = j$ given previous state $L_{t-1} = i$, i.e., $a_{ij} = P(L_t = j|L_{t-1} = i)$. 
In their simplest form, these probabilities are constants across time, but they may also be defined as functions of observed time-varying covariates. 
The last element of a hidden Markov model describes the distribution or probability of the response variable given the current state, $b_{j}(y_t) = P(Y_t = y_t|L_t = j)$. 

The auto-regressive hidden Markov model (ARHMM) adapts the HMM to a context where the Markovian assumption that $Y_t \perp\kern-5pt\perp Y_{t-k} | L_t$ for any $k\geq 1$ is violated \cite{juang1985mixture, ephraim2005revisiting}. It allows for $Y_t$ to depend on up to $k$ lagged values of the response, meaning for integer $k>1$, $Y_t\perp\kern-5pt\perp Y_{t-k-1} | L_t, Y_{t-1},..., Y_{t-k}$. Thus in an ARHMM, the response distribution will depend not only on the current latent state but also on lagged values of the response variable, meaning $b_{j}(y_t) = P(Y_t|L_t = j, Y_{t-1},...Y_{t-k})$. 
Under a frequentist approach, the initial state probabilities, transition probability matrix, and response distribution are often identified using the Baum Welch algorithm, a forward and backward expectation maximization approach \cite{baum1970maximization}. One first calculates recursively for each $i = 1,...,J$ the forward and backward probabilities, namely $\alpha_i(t) = P(Y_1 = y_1, ..., Y_t = y_t, L_t = i)$ and $\beta_i(t) = P(Y_{t+1} = y_{t+1}, ..., Y_T = y_T| L_t = i)$, respectively. These probabilities are then used to update the parameters, and the process is repeated until convergence is reached. Following convergence, researchers commonly employ the Viterbi algorithm to recursively backwards identify the most plausible sequence of latent states given the parameter estimates \cite{forney1973viterbi}. 

\section{Model specification}
\label{s:our model}

As seen in Figure \ref{fig p2 dag}, the hypothesized causal structure includes complex relationships and feedback loops among time-varying observed and unobserved variables. Thus the existing hidden Markov model literature requires numerous adaptations for such a model to be appropriate for estimating the causal relationships of interest. 
First, the model must include three sets of dependent variables to be modeled as descendants of $L_t$: $Y_t, A_t,$ and $C_t$. Each of these time series will require distributional assumptions and parameters to be estimated within the model. We will denote the relative probability distributions given $L_t = j$ as $b_{j,Y_t}(y_t), b_{j,A_t}(a_t), b_{j,C_t}(c_t)$ for $Y, A,$ and $C$, respectively. We note that within the hidden Markov model literature, these three variables would be labeled as response variables, however we refrain from this terminology, as in our causal framework only $Y_t$ is our response of interest, while $A_t$ is the intervention and $C_t$ a covariate. 

The HMM elements introduced in Section \ref{s:our model} must be defined as functions dependent on covariates of interest. We assume that the initial probabilities are independent of all observed and unobserved variables. However, from Figure \ref{fig p2 dag} we see that for $t>1$, $L_t$ depends on not only $L_{t-1}$ but also $C_{t-1}, W_t, Y_{t-1},$ and  $A_{t-1}$. Thus the transition probability matrix $\textbf{A}_t$ must be defined uniquely at each time point $t$ such that each element is dependent on the causal parents of $L_t$. If $J = 2$, we use a logit link, and for $J > 2$ we use a multinomial logit link. We let 

\begin{align}  
\eta_{L_t, ij} = \lambda_{1,ij} + \lambda_{2,ij}C_{t-1}+ \lambda_{3, ij}W_{t-1}+ \lambda_{4,ij}Y_{t-1} + \lambda_{5,ij}A_{t-1}.
\end{align}

\noindent Note that $\eta_{L_t, ij}$ and $\lambda_{ij}$ only need to be identified for $j>1$, as for $j= 1$ the probability will be implied using the link function. Elements $b_{j,Y_t}(y_t), b_{j,A_t}(a_t), b_{j,C_t}(c_t)$ must also be estimated not only dependent on $L_t$, but also their relative ancestors. Thus, we have: 

\begin{align}
E(Y_t|L_t = j, Y_{t-1},A_t, C_t, W_t, A_{t-1})  
&= f^{-1}(\beta_{1,j} + \beta_{2} Y_{t-1}+ \beta_{3,j} A_t + \beta_{4} C_t + \beta_{5} W_t +  \beta_{6,j} A_{t-1});\\
E(A_t|L_t = j, A_{t-1}, W_t, C_{t-1}, Y_{t-1})&= 
g^{-1}(\psi_{1,j} + \psi_{2} A_{t-1}+ \psi_{3} W_t + \psi_{4} C_{t-1} + \psi_{5} Y_{t-1});\\
E(C_t|L_t = j, C_{t-1}, W_t, A_{t-1}, Y_{t-1})  
&= h^{-1}(\phi_{1,j} + \phi_{2} C_{t-1}+ \phi_{3} W_t + \phi_{4} A_{t-1} + \phi_{5} Y_{t-1})
\end{align}

\noindent where $f(), g(),$ and $h()$ represent the link function dependent on the assumed distribution of $Y_t, A_t$, and $C_t$, respectively. Note we assume $L_t$ only modifies the relationship of $A_t$ on $Y_t$, however this could be relaxed. 
From the above expressions and distributional assumptions, $b_{j,Y_t}(y_t), b_{j,A_t}(a_t)$ and $b_{j,C_t}(c_t)$ can be easily obtained by evaluating the probability or density of a given observation. We also calculate the joint probability as $b_{j,Y_t,A_t,C_t}(y_t, a_t, c_t) = b_{j,Y_t}(y_t) b_{j,A_t}(a_t) b_{j,C_t}(c_t)$ to represent the plausibility of the descendants of $L_{t}$ given $L_{t} = j$ and other historical information. 

\section{Model implementation}
\label{s: mod_imp}

We consider a frequentist hidden Markov model approach (HMM-F) to estimate the parameters defined in Section \ref{s:our model}. In Appendix \ref{s:append_Sim}, we explore an equivalent model fit within a Bayesian framework using a Markov Chain Monte Carlo sampling procedure. For the frequentist model, we implement an expectation-maximization (EM) procedure based on the Baum-Welch algorithm. In the expectation step,  forward probabilities $\alpha_{i}(t)$ and backward probabilities $\beta_i(t)$ are calculated recursively using $a_{ij,t-1}$ and $b_{j,Y_{t},A_{t},C_{t}}(y_{t}, a_{t}, c_{t})$ \cite{baum1970maximization, juang1985mixture} such that

\begin{align}
    \alpha_{i}(t) &=  b_{j,Y_{t},A_{t},C_{t}}(y_{t}, a_{t}, c_{t})\sum_{j = 1}^M\alpha_j(t-1)a_{ij,t-1}; \\ 
    \beta_i(t) &= \sum_{j = 1}^M\beta_j(t+1)a_{ij,t}b_{j,Y_{t+1},A_{t+1},C_{t+1}}(y_{t+1}, a_{t+1}, c_{t+1}).
\end{align}

We then use the forward and backward probabilities to calculate weights to be used in the maximization step updating parameters. Specifically, we obtain: 

\begin{align}
    \gamma_i(t) &= P(L_{t} = i|Y, A, C, W) = \frac{\alpha_i(t)\beta_i(t)}{\sum_{j = 1}^M\alpha_j(t)\beta_j(t)};\\
    \xi_{ij}(t) &= P(L_t = i, L_{t+1}= j|Y, A, C, W) = \frac{\alpha_i(t)a_{ij,t}\beta_j(t+1)b_j(y_{t+1},a_{t+1}, c_{t+1})}{\sum_{k= 1}^M\sum_{w=1}^M\alpha_k(t)a_{kw,i}\beta_w(t+1)b_w(y_{t+1},a_{t+1}, c_{t+1})}.
\end{align}
To update the $\beta$, $\psi$, and $\phi$ estimates, we use the appropriate generalized linear regression for $Y_t$, $A_t$, and $C_t$ as the outcome with $\gamma_i(t)$ specified as the weights. To update the $\lambda$ estimates, we use logistic regression if $L$ is binary and a multinomial log-linear model \cite{ripley2016package} if $J>2$, with $\xi_{ij} (t)$ used as weights. The EM process is repeated until convergence is reached. Once convergence has been attained, we use the Viterbi algorithm to predict the most likely sequence for $L_t$ given the estimated parameters and observed data. 

\subsection{Missing data}

As it is likely that the outcome variable is subject to missingness, we propose a method of handling missing data within the auto-regressive hidden Markov model. To account for the loss of information, we marginalize over any missing $Y_t$ when it is treated as the response, meaning if $Y_t$ is missing, $b_{j,Y_t}(y_t) = 1$. However, to avoid being overly burdened by missing data we use imputation when $Y_t$ is to be treated as a regressor \cite{cai2022state}. For example, $Y_t$ is used as a lagged regressor for $Y_{t+1}$ and a regressor for $L_{t+1}$. In such instances, we use an imputation $\hat{y_t}$. 
We first consider a hidden Markov model under the frequentist implementation using a singular imputation (HMM-F-S), where at the end of each maximization step, we singularly and recursively across time impute $\hat{y_t} = \sum_{i = 1}^J \gamma_{i}(t)E(Y_t|L_{t}=i, A_t, C_t, W_t, A_{t-1})$. This approach represents incorporating the missingness in $Y$ into the overall expectation-maximization (EM) algorithm, as the imputations of $Y_t$ are set as the expectation using the current estimated coefficients. 

We additionally consider a hidden Markov model under the frequentist implementation which employs multiple imputation. For this approach, for imputation $m$ we sample $\hat y_t ^{(m)}$ from its assumed distribution, with mean $\sum_{i = 1}^J \gamma_{i}(t)E(Y_t|L_{t}=i, A_t, C_t, W_t, A_{t-1})$. While traditionally in an HMM, $L_t$ is treated probabilistically with weights to represent the plausibility of different states at each time point, here we consider additionally multiply imputing $L_t$. This imputation occurs after the calculation of forward and backward probabilities, at which point we impute $L_t$ recursively from $P(L_t=j|L_{t-1}=i, \hat{Y}^{(m)}, A,C,W) = \gamma_j(t)/ \xi_{ij}(t-1)$. The linear regressions are then fit using the imputations for $L_t$, with no weights. We denote this method as HMM-F-ML (hidden Markov model, frequentist, multiple imputation of latent state). Finally, at the end of the M-step, we first pool the estimates using Rubin's rules \cite{little2019statistical}, and then calculate new imputations for $Y_t$ using the pooled estimates. For both the singular and multiple imputation, we use Kalman smoothing imputation to initialize. We note that in Appendix \ref{s:append_Sim} we also consider a hybrid of these two approaches, where $Y_t$ is multiply imputed but $L_t$ is treated probabilistically with weights to represent the plausibility of different states at each time point (HMM-F-M).


 

\subsection{Inference}

We employ a block bootstrap approach to obtain confidence intervals for the frequensist implementation, as is common in time series analysis literature \cite{buhlmann2002bootstraps, hardle2003bootstrap}. We specifically follow the block bootstrap procedure proposed by Cl{\'e}men{\c{c}}on and Tressou \cite{clemenccon2009pseudo} for a hidden Markov model, which accounts for discontinuities that might be induced by separating blocks at random points when an underlying latent state is present. This method recommends first fitting the HMM model, obtaining a predicted sequence of the latent state $L_t$, and using this sequence to generate blocks to be sampled with replacement. Specifically, the observed data is split at each time point the predicted $L_t$ returns to the most frequently occurring state from a different state at $L_{t-1}$. This procedure is used to avoid non-stationary disruptions that might be introduced by simpler block sampling algorithms. For each bootstrap, the blocks are then sampled with replacement based on probabilistic weights calculated by the block length until a new time series is obtained with a length at least as long as the original data. The model is re-fit on each resulting multivariate time series of observed variables, and we use a percentile confidence interval from the bootstrap estimates. We denote this bootstrap with `B' for block bootstrap (HMM-F-S-B, HMM-F-M-B, or HMM-F-ML-B dependent on the imputation procedure). 

In Appendix \ref{s:append_Sim} we additionally consider a parametric bootstrap as was proposed by Visser et al. \cite{visser2000confidence}. This approach, following the convergence of the HMM model, simulates a new multivariate time series of the observed variables using the obtained estimates and re-fits the HMM model for each bootstrap. We then obtain a percentile confidence interval from the bootstrap estimates. We denote this method as HMM-F-S-P for singular imputation of $Y_t$, HMM-F-M-P for the multiple imputation of $Y_t$, and HMM-F-ML-P for the multiple imputation of $Y_t$ and $L_t$, where the `P' denotes parametric bootstrap. 
In Figure \ref{fig:flowchart} we visually display the differences in the six proposed frequentist algorithms.



\section{Simulation}
\label{s:Sim}

We consider two simulation setting with $J=2$ or $J=3$ disease states, and $T=500$ follow-up points. We include two internal covariates $C_t = \{C_{1,t}, C_{2,t}\}$ and two external covariates $W_t = \{W_{1,t}, W_{2,t}\}$. We simulate $Y_t, C_{1,t}, C_{2,t}$, and $W_{1,t}$ as continuous with normally distributed errors. We let $W_{2,t}$ represent the indicator of it being a weekend night. For each setting we generate 200 simulations. We focus on a scenario which considers an $A_t$ as continuous with normal errors, however in Appendix \ref{s:append_Sim} we explore the performance of the proposed methods in an additional scenario where $A_t$ is generated and modeled as an ordinal variable, with proportional odds of ascending to an additional state of $A_t$ assumed to be constant. We simulate $Y$ to be approximately 30\% missing at random, and all other variables are fully observed. To each resulting multivariate time series, we first apply a naive linear model which assumes stationarity and does not identify nor account for modification by $L_t$. For this model, we obtain a 95\% confidence interval using traditional asymptotic assumptions of normality. For the proposed methods, we complete 1000 bootstraps for each model type and calculate the empirical 95\% confidence interval from the bootstraps. For the multiple imputation model, HMM-F-ML-B, we include ten multiple imputations. 
For the two proposed hidden Markov models, after convergence is reached we use the Viterbi algorithm to estimate the most likely sequence of $L_t$ and compare this to the true $L_t$ used to simulate the data. We then re-label the imputed states to maximize the possible prediction accuracy. 
This procedure addresses the potential that a model may switch the labels for the latent states, but still be accurate in prediction and estimating coefficients of interest. 

\begin{figure}
\centerline{%
\includegraphics[width = 150mm]{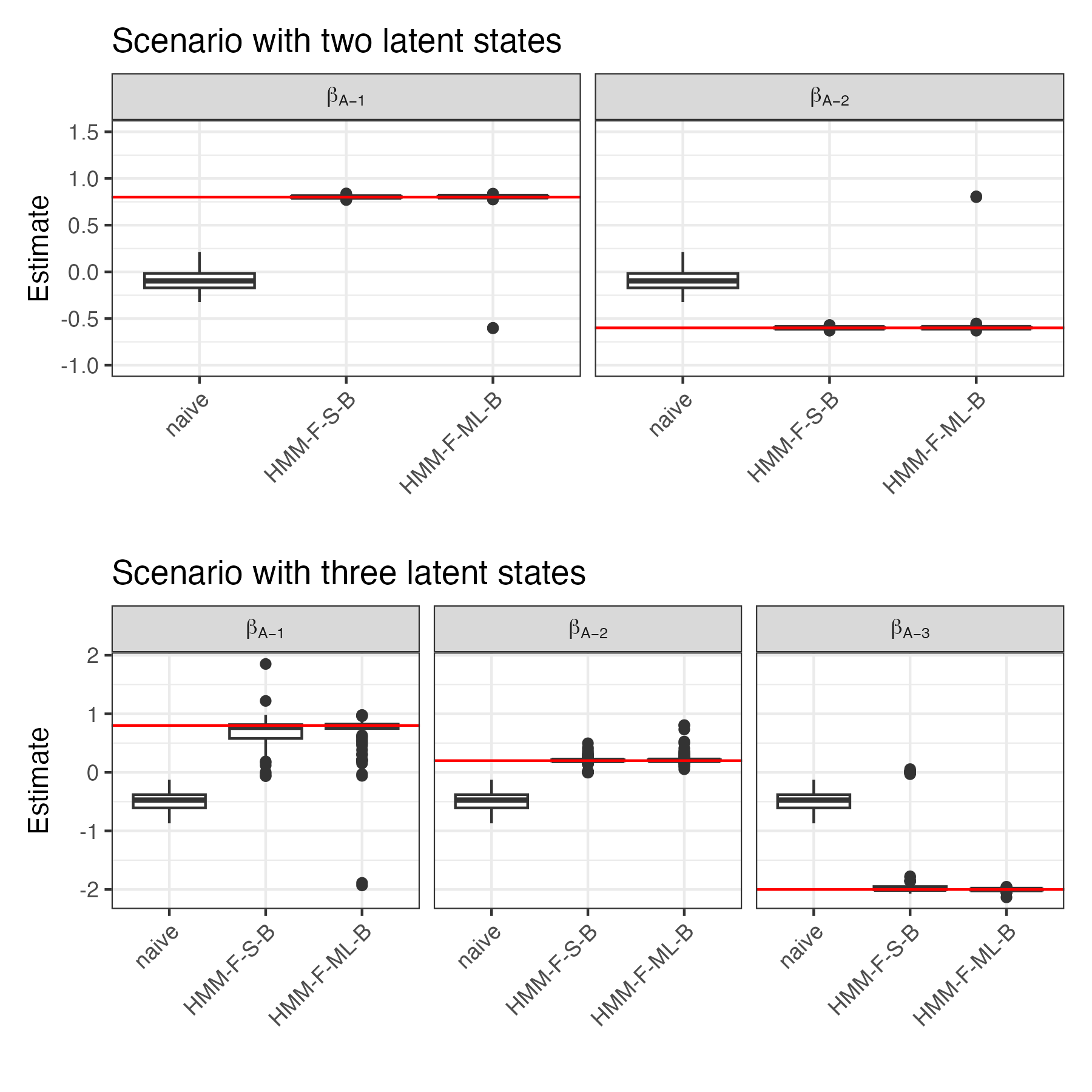}}
\caption{Simulation results for the naive model, hidden Markov model frequentist, singular imputation with block bootstrap (HMM-F-S-B); and hidden Markov model frequentist, multiple imputation of $L$ and $Y$, with block bootstrap (HMM-F-ML-B). Each row displays box plots of each method's estimates of the effect of $A_t$ on $Y_t$ with $L_t = 1$, $L_t = 2$, and $L_t = 3$, with the truth shown as a horizontal red line, with the first row showing results for the scenario with two latent states and the second row showing results for the scenario with three latent states.} 
\label{fig:sim_estimates}
\end{figure}

\begin{figure}
\centerline{%
\includegraphics[width = 150mm]{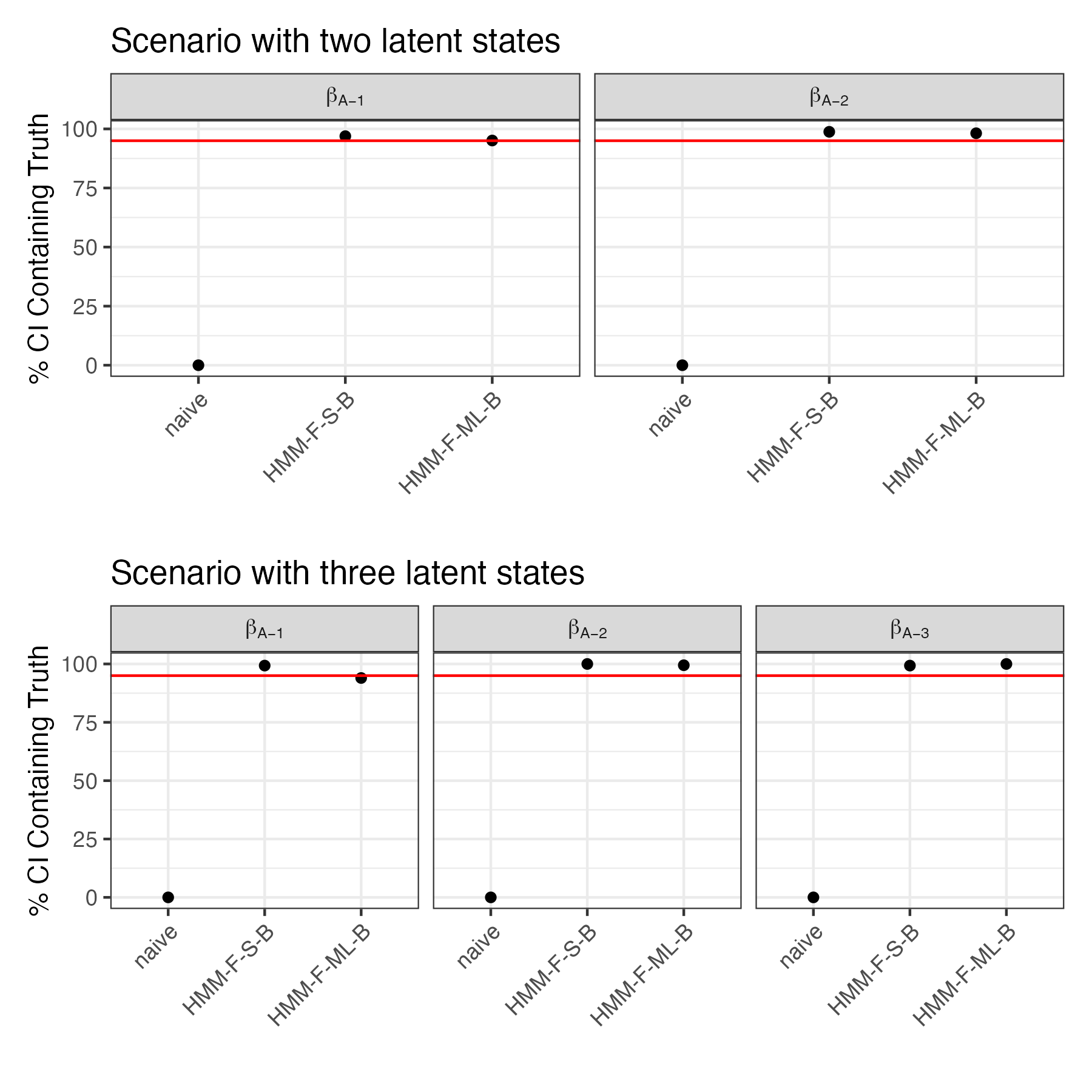}}
\caption{Simulation results for the naive model, hidden Markov model frequentist, singular imputation with block bootstrap (HMM-F-S-B); and hidden Markov model frequentist, multiple imputation of $L$ and $Y$, with block bootstrap (HMM-F-ML-B).  Each row displays the percent of 95\% confidence intervals that contain the truth, with 95\% indicated by the red horizontal line for each method's estimates of the effect of $A_t$ on $Y_t$ with $L_t = 1$, $L_t = 2$, and $L_t = 3$, with the first row showing results for the scenario with two latent states and the second row showing results for the scenario with three latent states.} 
\label{fig:sim_coverage}
\end{figure}

In Figure \ref{fig:sim_estimates}, we plot the results for the estimates of $\beta_{3,1}$, $\beta_{3,2}$, and $\beta_{3,3}$, the contemporaneous conditional average individual causal
effect of $A_t = a_{t} + 1$ versus $A_t = a_{t}$ on $Y_t$ given $L_t = 1$, $L_t = 2$, and $L_t = 3$, respectively. The first row displays results from the scenario with two underlying latent states (only $\beta_{3,1}$ and $\beta_{3,2}$ estimated), while the second shows results with a multinomial $L_t$ with three underlying latent states. Most notably here, we see how the naive approach fails to estimate the effects of interest, as it does not account for modification by the latent state.
We note that the estimates for both proposed models are very accurate and unbiased. 
In Figure \ref{fig:sim_coverage}, we show the confidence interval coverage, representing the percent of the 95\% CI intervals which include the truth across simulations. As might be expected given the bias in the estimates for the naive approach, this method achieves 0\% coverage. However, both proposed methods achieve acceptable coverage regardless of whether $L_t$ is binary or multinomial. 
While the estimates for the coefficient of $A$'s effect on $Y$ across the HMM models are consistently unbiased in the simulation, we note that several other estimates do tend to be slightly biased, likely due to multicollinearity.

In Table \ref{tab:pred_acc} we provide the prediction accuracy of predictions of latent state from the Viterbi algorithm compared to the true underlying latent state.
We note that as the naive model does not identify the latent state, it has no such accuracy. The best performing model in terms of prediction accuracy is the frequentist implementation with single imputation, HMM-F-S-B, with accuracy of 97.3\% for two latent states, and 90\% for three latent states. 
The frequentist implementation with multiple imputation, HMM-F-ML-B,  performs worst for prediction accuracy, with only 85.7\% accuracy in two latent state scenario, and 73.6\% accuracy in the three latent state scenario. 
Finally, we note that in Appendix \ref{s:append_Sim}, we provide additional simulation results for a hybrid frequentist model that multiply imputes $Y_t$ but treats $L_t$ probabilistically, and frequentist models using a parametric bootstrap, as well as results of the proposed methods in a simulation setting with an ordinal exposure $A_t$. 

\begin{table}

\caption{Prediction accuracy across settings with two or three underlying latent states in simulations with a continuous exposure $A_t$ with normally distributed errors for the hidden Markov model frequentist, singular imputation with block bootstrap (HMM-F-S-B); hidden Markov model frequentist, multiple imputation of $L$ and $Y$, with block bootstrap (HMM-F-ML-B). Note the naive model does not identify latent states and thus has no prediction accuracy.}
\label{tab:pred_acc}
\begin{center}

\begin{tabular}[t]{l c c}
\toprule
\textbf{Model} & \hspace{8mm}\textbf{Two latent states}\hspace{8mm} & \hspace{8mm}\textbf{Three latent states}\hspace{8mm}\quad\\
\hline
naive & - & -\\
HMM-F-S-B & 0.973 & 0.900\\
HMM-F-ML-B & 0.857 & 0.736\\

\toprule
\end{tabular}

\end{center}

\end{table}

\section{Application}
\label{s:app}

The Bipolar Longitudinal Study (BLS) is an ongoing mobile health study at McLean Hospital where 74 patients with schizophrenia or bipolar disorder are followed for up to five years. From the individual's smartphone, their anonymized call and text logs, GPS movements, accelerometer behavior, and daily symptom, emotion, and behavior survey responses are recorded via the Beiwe application \cite{onnela2021beiwe}. The patients additionally participate in approximately monthly in-person visits where a range of assessments are conducted. Here, we consider a several distinct exposures of interest which are captured using both active ecological momentary assessment (EMA) data, and passive data to show how the proposed approach can allow for the integration of different data modalities.

Here we first consider a female participant with a bipolar spectrum disorder and a follow-up of 708 days. Our outcome of interest ($Y$) is overnight sleep duration, as identified from watch actigraph data via the deep phenotyping of sleep (DPSleep) algorithm \cite{rahimi2021open}.  We use daily total incoming and outgoing text counts with key contacts as the exposure, where key contacts were identified as individuals with whom the participant is in frequent reciprocated contact. To account for skewness, in the model we use a square root transformation on the exposure, meaning $A_t = \sqrt{\text{text count with key contacts}}$. We additionally include negative and positive mood in the model as covariates, $C_{1,t}, C_{2,t}$, respectively. Negative and positive mood are calculated as an aggregate score from 0 to 28 of negative and positive items on the daily survey. Lastly, we use the local daily max environmental temperature as $W_{1,t}$, and whether the night is a weekday or weekend as $W_{2,t}$, covariates that are not affected by the participant. We apply our proposed modeling methods to the individual's data, using the same specifications as in the simulation. We specify the model to include two underlying latent states, as clinically we understand this patient only exhibits symptoms related to depressed and euthymic states. 
We note that while $A_t$ and $W_t$ are fully observed, 23\% of $C_t$ is missing, representing non-response for the daily EMA data. As the model currently is only designed to allow missing data in $Y$, we impute missing values in $C$ using Kalman smoothing imputation prior to fitting the model. Overnight sleep duration, $Y_t$ is missing 20\% of the days across follow-up, and we leave this missingness to be handled by the models. 

\begin{figure}
\centerline{%
\includegraphics[width = 170mm]{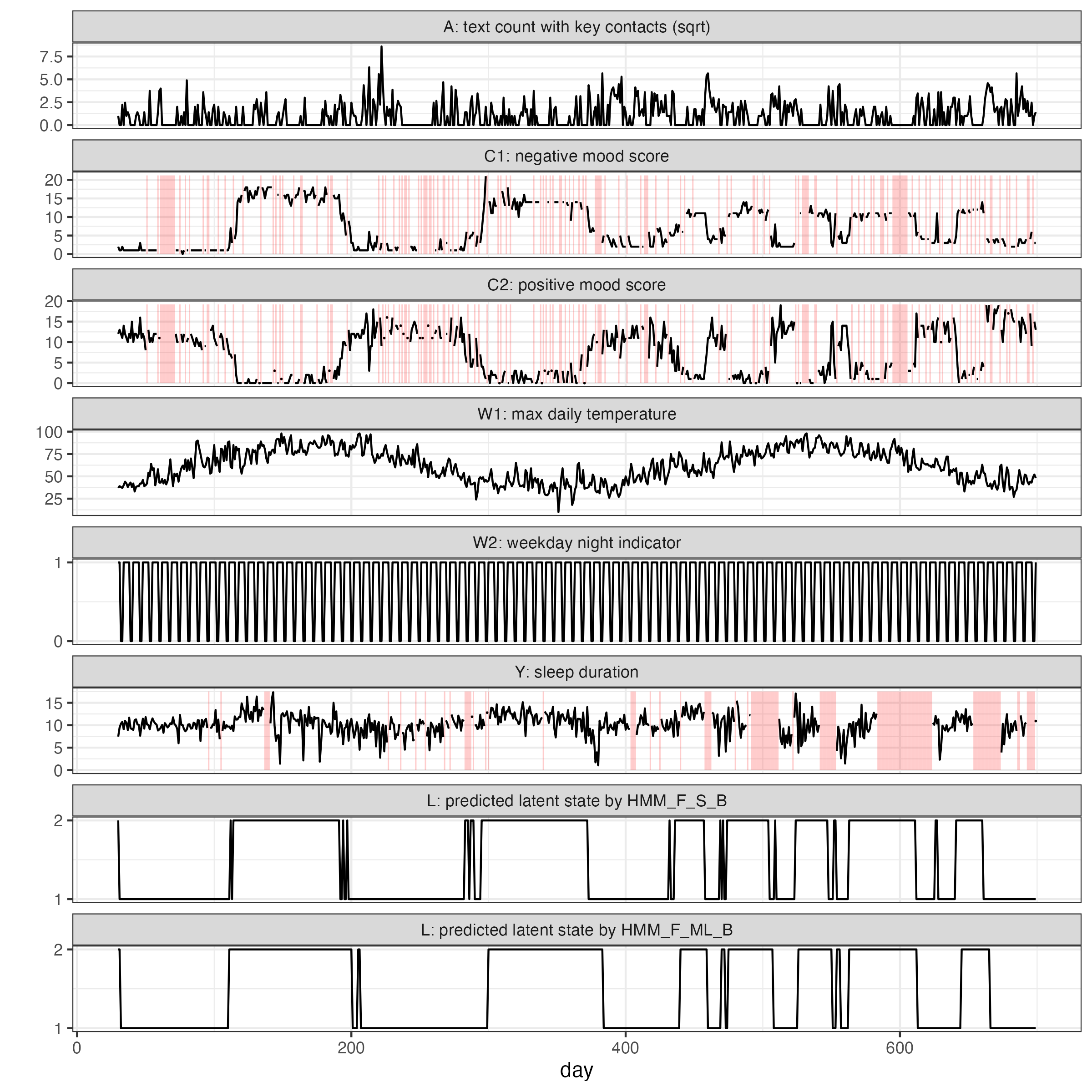}}
\caption{Multivariate time series of Paricipant A with bipolar spectrum disorder from the Bipolar Longitudinal Study. The first six rows show observed characteristics, with missing values highlighted in red, while the fifth and sixth rows plot the predicted disease states obtained from the HMM-F-S-B and HMM-F-ML-B models, respectively. } 
\label{fig:5bt_p2}
\end{figure}

In Figure \ref{fig:5bt_p2}, we show the time series for Participant A of the six variables included in the analysis as well as the predicted latent disease state from the frequentist models using single and multiple imputation: HMM-F-S-B and HMM-F-ML-B, respectively. 
From the observed information in rows 1-6, and particularly the mood scores, we can visually identify how the patient cycles in and out of a depressive state. In rows 7-8, we see how both frequentist approaches are largely able to identify the individual's disease state. Notably, the predicted states appear to most fluctuate around the times when the individual is shifting from a depressive to euthymic or euthymic to a depressive state. This reflects a limitation, as while the model requires a given state to be assigned each day, in reality, it is more likely that this transitional period lasts a few days, where the individual is in between states. We find that the predictions of the latent state across the two models are fairly consistent, with the two models identifying the same latent state for 89.4\% percent of the days in the follow-up. Notably, the time points of the two latent states largely concord with the monthly in-person PANSS scores indicating the occurrence of a depressive state, confirming that the models are indeed identifying the underlying disease state. 

\begin{table}

\caption{Estimates and confidence intervals for the contemporaneous conditional average individual causal effect, and 1-day and 2-day lagged conditional average individual causal effect of digital activity on sleep duration under the first and second latent states, resulting from the frequentist model with singular imputation (HMM-F-S-B) and multiple imputation (HMM-F-ML-B) for Participant A.}
\label{tab:p2 5bt est}
\begin{center}
\begin{tabular}[t]{l l l l l l}
\toprule
 & & \textbf{effect} & \textbf{estimate} & \textbf{95\% CI low} & \textbf{95\% CI high}\\
\hline
\textbf{HMM-F-S-B} & $\mathbf{L_t = 1}$ & contemporaneous & 0.031 & -0.199 & 0.089\\
 &  & 1-lag & -0.088 & -0.121 & 0.001\\
 &  & 2-lag & -0.075 & -0.051 & 0.055\\
 & $\mathbf{L_t = 2}$ & contemporaneous & 0.14 & -0.223 & 0.036\\
 &  & 1-lag & -0.011 & -0.098 & 0.021\\
 &  & 2-lag & -0.023 & -0.017 & 0.081\\
\hline
\textbf{HMM-F-ML-B} & $\mathbf{L_t = 1}$ & contemporaneous & 0.047 & -0.164 & 0.135\\
 &  & 1-lag & -0.064 & -0.148 & 0.075\\
 &  & 2-lag & 0.005 & -0.16 & 0.224\\
 & $\mathbf{L_t = 2}$ & contemporaneous & -0.038 & -0.277 & 0.402\\
 &  & 1-lag & -0.195 & -0.427 & 0.206\\
 &  & 2-lag & -0.211 & -0.329 & 0.406\\
\toprule
\end{tabular}
\end{center}

\end{table}

In Table \ref{tab:p2 5bt est} we provide the estimates and confidence intervals for the contemporaneous conditional average individual causal effect, and 1-day and 2-day lagged conditional average individual causal effect of digital activity on sleep duration under the first and second latent states, resulting from the two frequentist models for Participant A. We find no significant effects between digital socialization and sleep duration for the contemporaneous, 1-lag, and 2-lag effects, regardless of the latent disease state. Interestingly, it appears that the point estimates for the contemporaneous effect tend to be positive, but the lagged effect point estimates tend to be negative, indicating that perhaps an increase in digital socialization increases sleep that night, but then results in reduced sleep for subsequent nights. 

In Figure \ref{fig:8RC_p2} we plot the observed time series and model results for a male individual, Participant B, with bipolar spectrum disorder and a follow-up of 984 days. Here, while we include the same outcome and covariates as for Participant A, we define the exposure of interest using GPS. The exposure $A_t$ specifically denotes the percentage of time spent away from home on a given day, which is calculated using the DPLocate pipeline 
\cite{rahimi2022measures}. Notably this individual has a much higher rate of missing data, with 68\% of EMA responses unobserved. Unfortunately, likely due to the high rate of missingness and low variability in the observed data, we see that the models are unable to capture clear trends for the latent disease state. While the frequentist model with single implementation (HMM-F-S-B) has rapid fluctuations between the two identified latent states, the model with multiple imputation (HMM-F-ML-B) has some more stable sections between periods with rapid fluctuations, but does not appear to capture meaningful disease states. Overall, this participant demonstrates a setting where the proposed methods are incapable of identifying a clinically relevant disease state, due to the high rate of missing data and lack of variability in the observed data. 

Lastly, we consider Participant C, a male individual with schizophrenia and depression, and a follow-up of 116 days. Here, we again use the same outcome of sleep duration in hours and covariates as with Participants A and B, but our exposure of interest is self-reported social activity in person. This item is collected on an ordinal scale from 0 to 4 with options: `I spent almost all of my time alone' (0); `I interacted with others but little more than superficial interactions' (1); `I interacted with others in a meaningful way' (2); `I extensively interacted with close friends or family or a significant other' (3); `I experienced an unusually deep connection with another person' (4). Notably the participant only responded with values ranging from 0 to 3. We treat this variable as ordinal using a proportional odds assumption. In Figure \ref{fig:5KX_p2} we plot the observed time series as well as the predicted latent states from the 
HMM-F-S-B and HMM-F-ML-B models. Similar to Participant B, neither model identifies clear latent disease state patterns. While this participant has much less missing data, their short follow-up with relatively little variability does not appear to provide sufficient information for the models to identify meaningful latent states. Overall, Participants B and C illustrate settings in which the proposed hidden Markov models are incapable of identifying an underlying disease state, because of insufficient recorded data and variability. Because of the poor fit for these two individuals, we do not report estimates of the effect of the exposure on the outcome conditional on the latent state, as the identified latent state does not have a meaningful interpretation. 




\section{Discussion}
\label{s:disc}

We introduce two hidden Markov models designed to perform causal estimation and inference while addressing latent modification, complex hypothesized causal relationships, and missing data in the outcome. We demonstrate in our simulation that the proposed models provide unbiased estimation of the contemporaneous average individual causal effect of interest and high prediction accuracy of the latent state. The simulation also demonstrates that confidence intervals resulting from the blocked bootstrap for the models provide adequate coverage. In addition to the two frequentist proposed models, we also explore the performance of an equivalent Bayesian model implemented using a Hamiltonian Monte Carlo algorithm through STAN \cite{carpenter2017stan} in Appendix \ref{s:append_Sim}. This model is similarly able to handle missingness in the outcome, and provides coverage intervals. In simulations, the prediction and estimation accuracy of the Bayesian approach is comparable to the two frequentist models, however, the approach fails to achieve sufficient coverage. Despite this limitation, we anticipate this method may be of interest in the future for sensitivity analyses, through the incorporation of informative priors. 

In an application to the Bipolar Longitudinal study, we see how the proposed frequentist hidden Markov models are able to recover a bipolar individual's disease states using mobile health data in the context where sufficient data and variability are observed, however, we fail to find significant results of the effect of digital socialization on sleep duration. We also identify in the application the limitations to the proposed methods when there is short follow-up or high rate of missing data. These limitations should be further explored through additional simulations. Overall, in the application we see that as long as sufficient data and variability between states is recorded, the model is able to capture and identify the underlying disease states of interest and make inferences.  

In general, we recommend the use of the single imputation approach (HMM-F-S-B) in cases with low rates of missing data, as the model is more computationally efficient, and achieves superior prediction accuracy in simulations. However in cases with higher rates of missingness, we recommend the model with multiple imputation (HMM-F-ML-B) as it is more robust at handling the increased uncertainty generated from the missingness. We note that while here we focus on applying the proposed methods to a singular participant at a time using an N-of-1 approach, we hypothesize the models could be extended to a hierarchical modeling where estimation and inferences are made not only at the individual level, but also across the sample of participants.

Our proposed methods provide an approach to estimate causal effects at the individual level in multivariate time series data, in the presence of latent modification, missing data, and complex causal relationships. This represents a substantial contribution to the latent modification literature, which primarily focuses on identifying latent subgroups of individuals who respond to treatment differently, rather than grouping of time points within time series data \cite{kim2015multilevel, kim2019estimating, shahn2017latent}. However, there are still several limitations to our approach. First, we must assume that stationarity holds after conditioning on the latent state, which may not be true for some individuals' behavior in mobile health studies, as seen in Fowler et al. \cite{fowler2024testing} and Cai et al\cite{cai2022state}. We additionally assume that the missingness is limited to the outcome variable, and that said missingness is ignorable. In truth, it is likely in mobile health that the missingness is not at random and across several variables, settings for which extensions of our proposed methods are needed. Another limitation is the limited resources on inference for HMMs. More work is needed to identify the ideal method for performing a bootstrap with a multivariate time series with an underlying latent state or to develop asymptotic inference results for complex hidden Markov models. We additionally note that our proposed methods are most accurate with low to moderate correlation between observed variables. The methods also suffer from underflow when many observed variables are specified as descendants of the latent state. Since mobile health studies may have numerous strongly autocorrelated time series, an extension that incorporates a shrinkage penalty is likely warranted. 

Lastly, more work is needed to comprehend the implications of a violation of the causal assumption for estimates. Specifically, the assumption of conditional exchangeability and thus no unobserved confounding likely does not hold for observational mobile health studies. Importantly, we note that given our hypothesized DAG, the latent state is a confounder for the relationship between the exposure and outcome. More work is needed to develop the necessary assumptions under which the effect of interest is still identifiable, given this latent confounding.

\section*{Funding}

This work is supported by National Institute of Mental Health K01MH118477 (PI: Valeri) and U01MH116925 (PI: Baker).

\section*{Acknoledgements}

We thank Dost Ongur for his valuable insights and thoughtful contributions.



\appendix
\section{Additional Simulation Results and Models}
\label{s:append_Sim}

This section provides supplementary simulation results including a simulation scenario with an ordinal exposure and two underlying latent states and the consideration of five additional hidden Markov models.

\begin{figure}
\centerline{%
\includegraphics[width = 150mm]{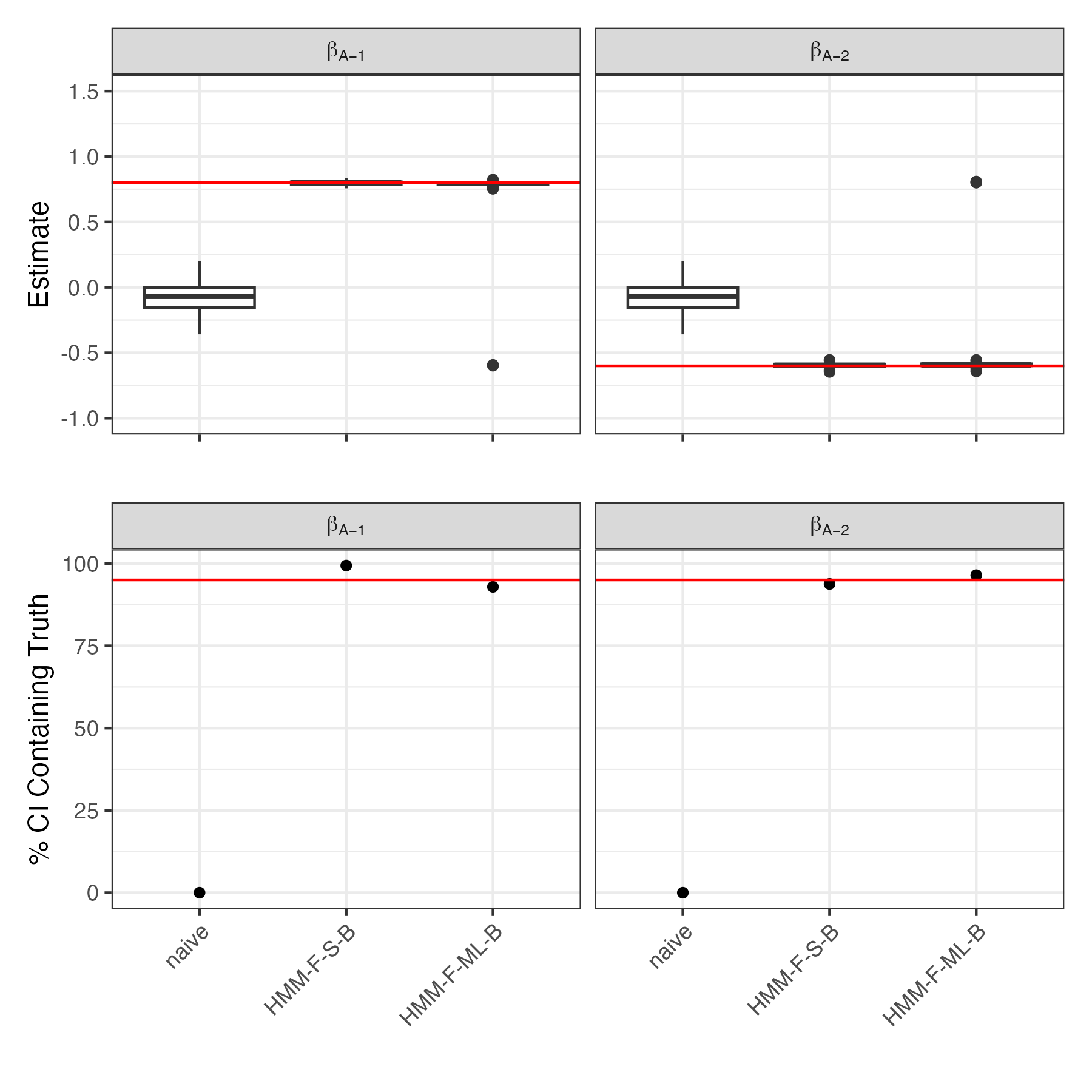}}
\caption{Simulation results for the naive model, hidden Markov model frequentist, singular imputation with block bootstrap (HMM-F-S-B); and hidden Markov model frequentist, multiple imputation of $L$ and $Y$, with block bootstrap (HMM-F-ML-B) with an ordinal exposure under two latent states. The first row displays box plots of
each method’s estimates of the effect of $A_t$ on $Y_t$ with $L_t = 1$, and $Lt = 2$, with the truth
shown as a horizontal red line. The second row displays the percent of 95\% confidence intervals that contain the truth, with 95\% indicated by the red horizontal line for each method's estimates of the effect of $A_t$ on $Y_t$ with $L_t = 1$, and $Lt = 2$.} 
\label{fig:sim_prop_A}
\end{figure}

In Figure \ref{fig:sim_prop_A} we plot the estimation and coverage results for the two proposed models in comparison to the naive model in a simulation setting with an ordinal exposure $A_t$, continuous outcome $Y_t$, and two underlying latent states. Notably, both proposed methods perform well in this scenario, achieving unbiased estimates. We note coverage is slightly lower for the multiple imputation approach, although still close to 95\%. In this simulation scenario, the approach with single imputation has a prediction accuracy of 0.985, while the multiple imputation approach has a prediction accuracy of 0.926.

In addition to the frequentist models proposed in Section \ref{s: mod_imp}, we consider a Bayesian implementation of the model described in Section \ref{s:our model} (HMM-B). The Bayesian procedure uses a forward algorithm to describe the latent state behavior, where for each time point $t$ and each possible state $i$, 
\begin{align}
    \alpha_j(t) = b_{j, Y_t, A_t, C_t}(y_t, a_t, c_t) \sum_{i=1}^J \alpha_{t-1}(i)a_{ij,t}.
\end{align}

The model is then specified to marginalize over the density of the log of the final states, i.e.  $\sum_{j = 1}^J log(\alpha_j(T))$. We optimize the model via a Hamiltonian Monte Carlo algorithm implemented in STAN \cite{carpenter2017stan}. To account for missing data in $Y$, we treat missing values of $Y_t$ as additional parameters within the model. We use diffuse non-informative priors for all parameters. To account for the latent states switching labels between samples of the Markov chain Monte Carlo, we apply the Viterbi algorithm to the parameter estimates and observed variables for each sampling and re-label the latent states to be consistent across samples. The estimates are then calculated as the mean of the sampling parameters, and a credible interval is obtained from the percentiles of the sampled estimates. A final prediction of $L_t$ is obtained by applying the Viterbi algorithm to the averaged estimates and observed data. 
 
\begin{figure}
\centerline{%
\includegraphics[width = 130mm]{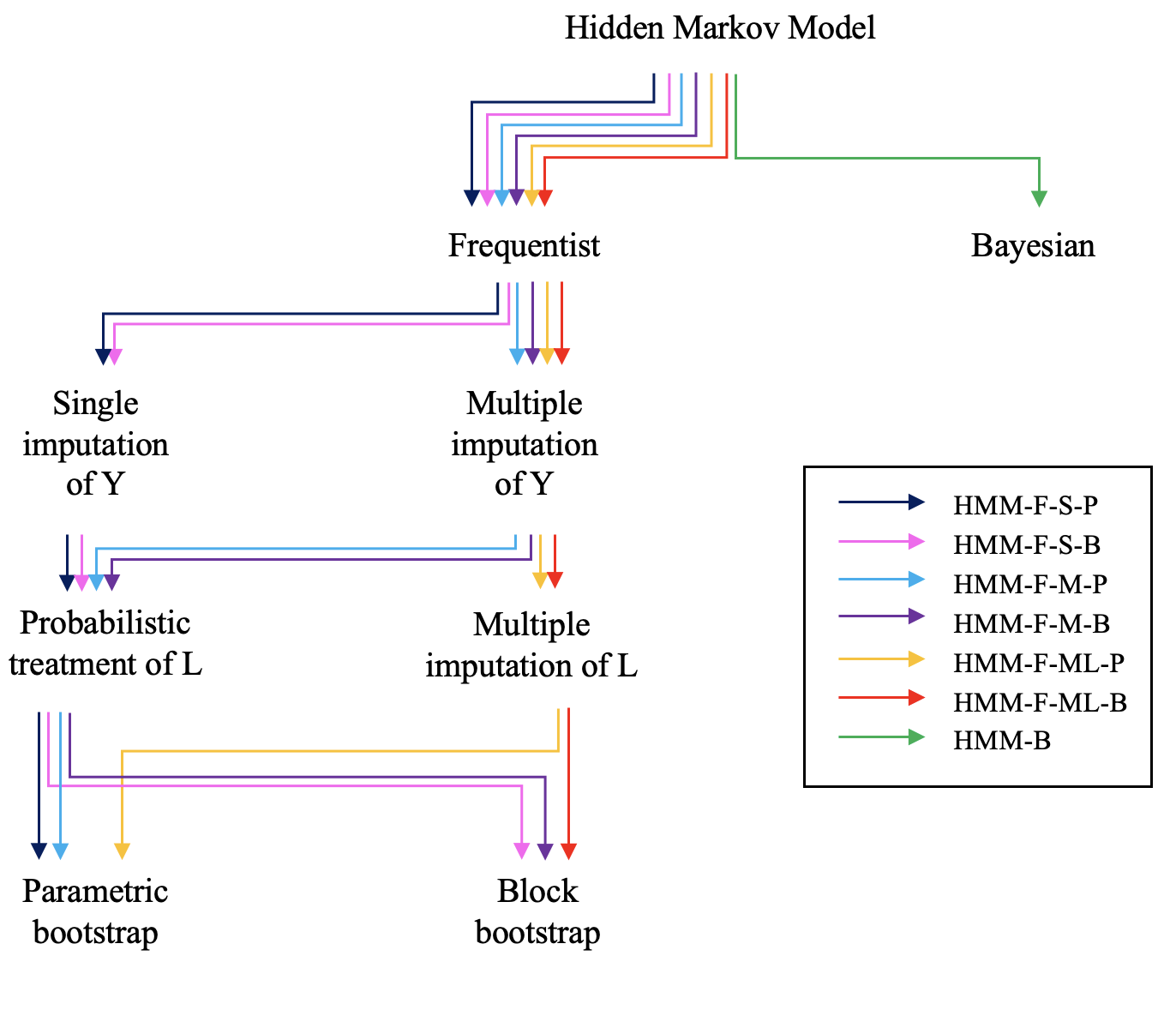}}
\caption{Flowchart of seven proposed HMM methods and their implementation and approaches towards missing 
in $Y$, latent state $L$, and bootstrapping. Note that the imputations of $Y$ are only used when $Y$ is to be treated as a regressor, and when $Y$ is missing as the outcome it is not imputed.} 
\label{fig:flowchart}
\end{figure}

\begin{figure}
\centerline{%
\includegraphics[width = 140mm]{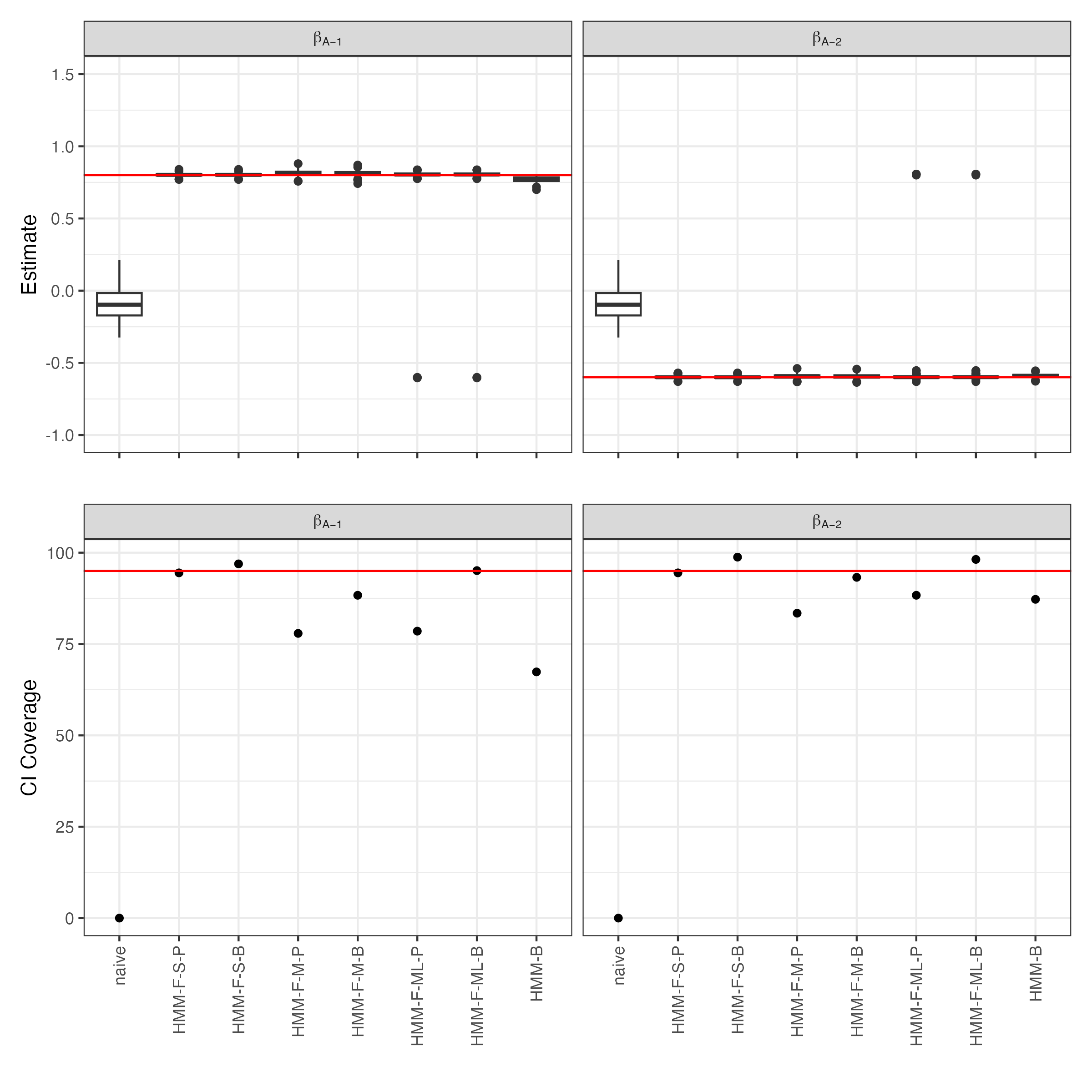}}
\caption{Simulation results for the naive model; hidden Markov models frequentist singular imputation parametric bootstrap (HMM-F-S-P) and block bootstrap (HMM-F-S-B); frequentist multiple imputation parametric bootstrap (HMM-F-M-P) and block bootstrap (HMM-F-M-B); frequentist multiple imputation of $L$ and $Y$ parametric bootstrap (HMM-F-ML-P) and block bootstrap (HMM-F-ML-B); and Bayesian (HMM-B). The first row displays box plots of each method's estimates of the effect of $A_t$ on $Y_t$ with $L_t = 1$ (left) and $L_t = 2$ (right), with the truth shown as a horizontal red line. The second row displays the percent of 95\% confidence/credible intervals which contain the truth, with 95\% indicated by the red horizontal line.} 
\label{fig:sim_normal_fullres}
\end{figure}

We also evaluate additional frequentist approaches. The first acts as a hybrid between the two algorithms described in Section \ref{s: mod_imp} where multiple imputation is used for the outcome, but the latent state is treated probabilistically (HMM-F-M). We also include an alternative procedure for inference using a parametric bootstrap, as described by Visser et al. \cite{visser2000confidence}, for each proposed frequentist estimation procedure. In Figure \ref{fig:flowchart} we visually show the differences between the different hidden Markov models. In Figure \ref{fig:sim_normal_fullres} we plot simulation results for the scenario with a continuous exposure and two latent classes across the seven proposed models and the naive approach. Overall, all seven models achieve relatively unbiased estimates. However, the hybrid approach with multiple imputation for the outcome but probabilistic treatment of the latent state (HMM-F-M), the Bayesian implementation (HMM-B) and the parametric bootstraps (HMM-F-S-P, HMM-F-ML-P) all suffer in the coverage of their 95\% confidence/credible intervals, indicating inferior performance for inference in comparison to the two models proposed in the main text (HMM-F-S-B and HMM-F-ML-B). The algorithm with multiple imputation of the outcome but probabilistic treatment of the latent state (HMM-F-M) has prediction accuracy of 89.2\% while the Bayesian approach (HMM-B) achieves a prediction accuracy of 97.6\%. We note the bootstrap method for the frequentist approaches does not impact the prediction accuracy of the models.

\newpage 
\section{Additional Participants}

\begin{figure}
\centerline{%
\includegraphics[width = 170mm]{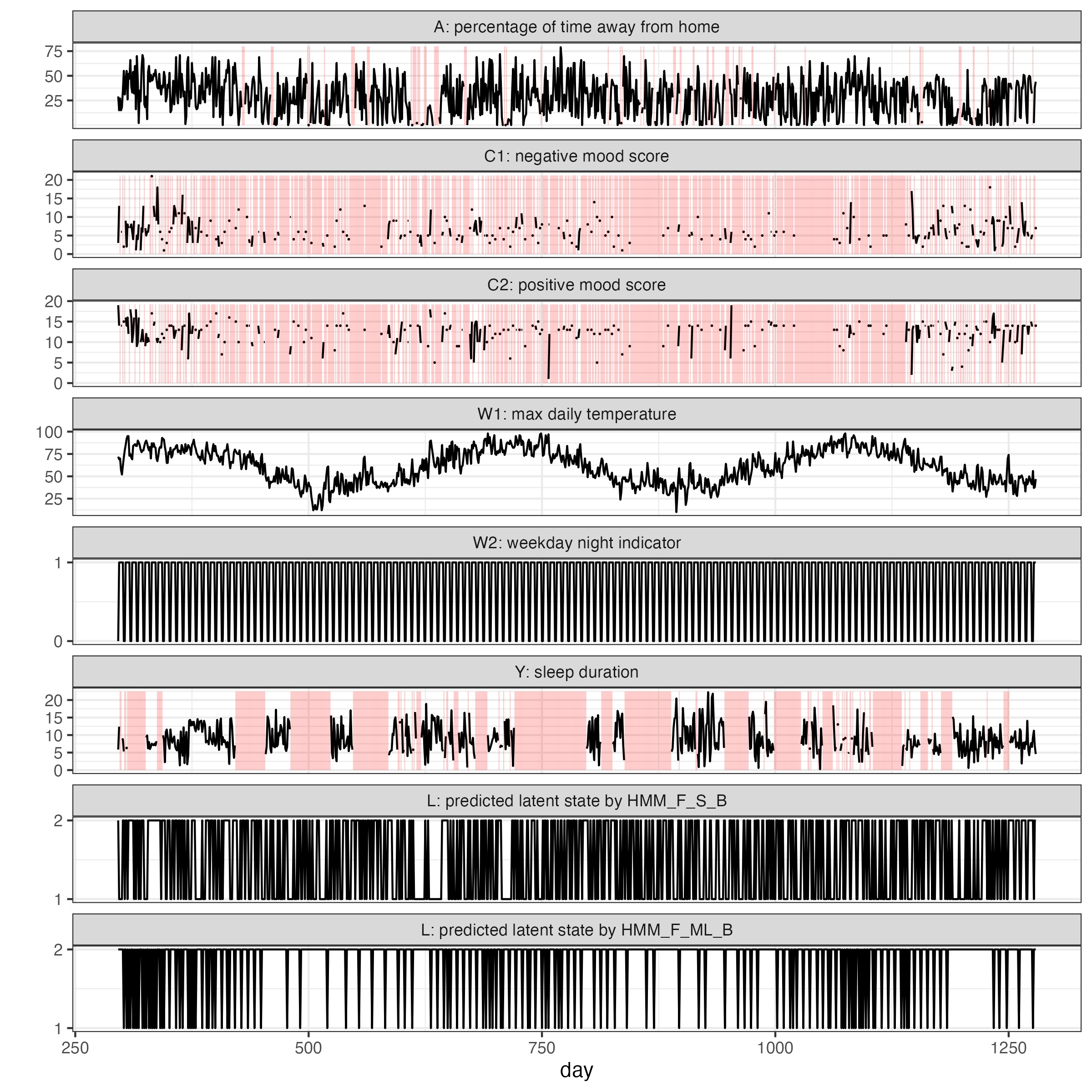}}
\caption{Multivariate time series of Participant B with bipolar spectrum disorder from the Bipolar Longitudinal Study. The first six rows show observed characteristics, with missing values highlighted in red, while the fifth and sixth rows plot the predicted disease states obtained from the HMM-F-S-B and HMM-F-ML-B models, respectively.} 
\label{fig:8RC_p2}
\end{figure}

\begin{figure}
\centerline{%
\includegraphics[width = 170mm]{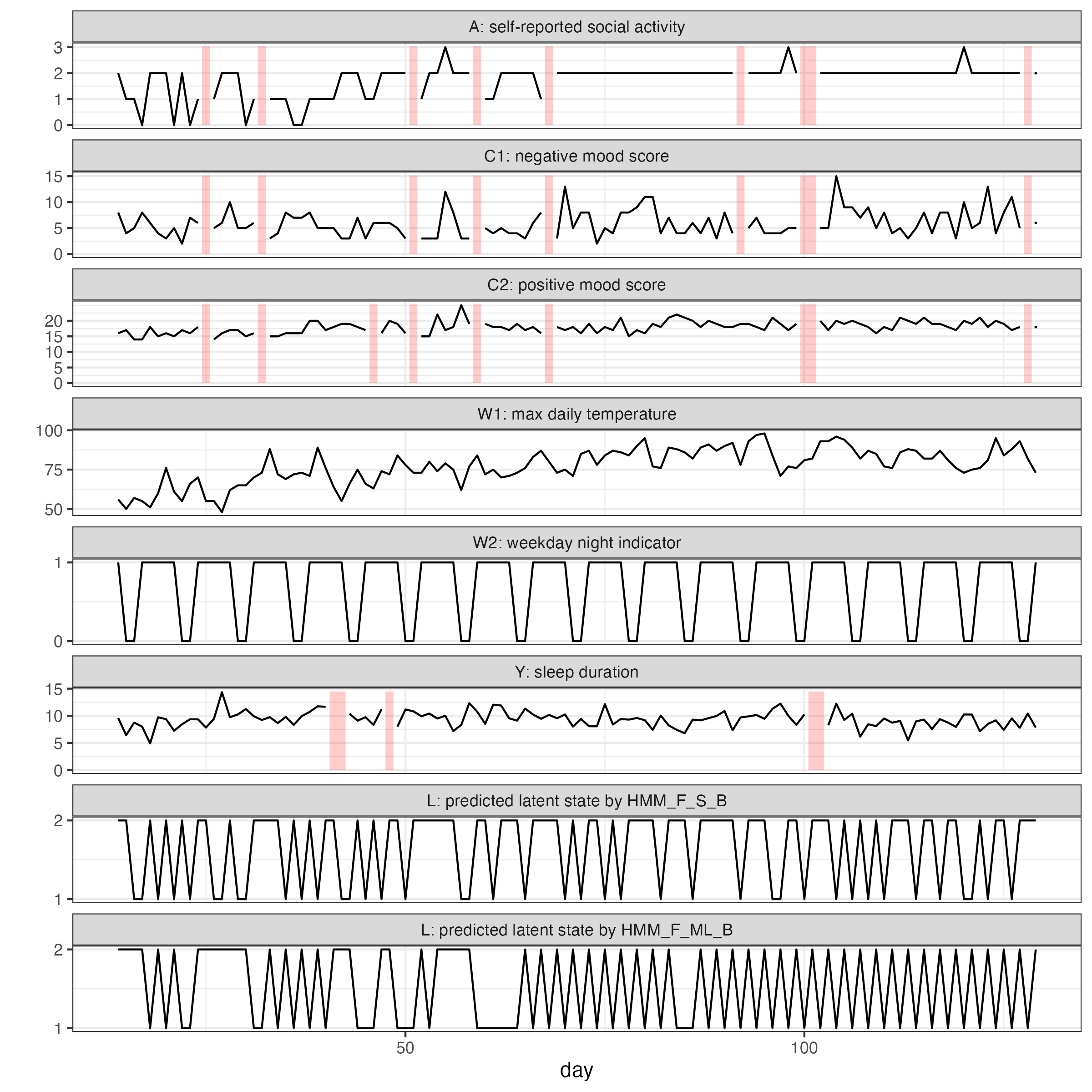}}
\caption{Multivariate time series of Participant C with bipolar spectrum disorder from the Bipolar Longitudinal Study. The first six rows show observed characteristics, with missing values highlighted in red, while the fifth and sixth rows plot the predicted disease states obtained from the HMM-F-S-B and HMM-F-ML-B models, respectively.} 
\label{fig:5KX_p2}
\end{figure}

This section provides supplementary figures of the observed data and the predicted latent state results for two additional participants from the Bipolar Longitudinal Study. In Figure \ref{fig:8RC_p2} we show the results of a male individual with bipolar spectrum disorder and 984 days of follow up, while in Figure \ref{fig:5KX_p2} we show results for a male individual with schizophrenia and depression and 116 days of follow up.  

\newpage 
\bibliographystyle{vancouver}
\bibliography{sampbib.bib}

\end{document}